\documentstyle[prl,aps]{revtex}

\input epsf
\tighten

\begin{document}
\title{A black hole solution to the cosmological monopole problem}

\author{Dejan Stojkovic and Katherine Freese}

\newcommand{\be}{\begin{equation}}
\newcommand{\ee}{\end{equation}}

\address{MCTP, Department of Physics, University of Michigan,  Ann Arbor,
 MI 48109-1120 USA}

\vspace{.8cm}

\wideabs{
\maketitle
\begin{abstract}
  \widetext We propose a solution to the cosmological monopole
  problem: Primordial black holes, produced in the early universe, can
  accrete magnetic monopoles before the relics dominate the energy density
of the universe. These small black holes  quickly
  evaporate and thereby convert most of the monopole energy density into
  radiation. We estimate the range of parameters for which
this solution is possible: under very conservative assumptions
we find that the black hole mass must be less than $10^9$gm.
\end{abstract}
 \hfill MCTP-04-22

}

\narrowtext

\bigskip

Magnetic monopoles are a generic prediction of Grand Unified
Theories (GUTs). Monopoles always appear as stable topological
entities in any GUT that breaks down to electromagnetism
\cite{tHooft,Polyakov}.  The monopole mass is $M_m \sim \eta/e$,
where $\eta$ is the energy scale of the phase transition and
$1/e=\sqrt{137}$. Hence, for $\eta \simeq 10^{15}$ GeV, monopoles of mass
$M_m \simeq 10^{16}$ GeV should exist. We use this mass scale as our
fiducial value
throughout, for the sake of definiteness, but note that an analogous
calculation can be done for any other scale\footnote{For example,
for $\eta \sim 10^{17}$GeV (which is favored by supersymmetric
extensions of the standard model and by experimental constraints
on the proton lifetime), the monopole mass would be $10^{18}$GeV.}.
The magnetic charge of the
monopoles, $g$, is determined by the quantization condition
$eg=n/2$ ($n$ is an integer). The abundance of these monopoles is
an open question. The Kibble mechanism \cite{Kibble} predicts
roughly one monopole per horizon volume at the time of the GUT
phase transition (due to causality constraints).  However, this
estimate provides a severe overabundance of the number of
monopoles: they overclose the Universe by many orders of
magnitude.  Since they are stable for topological reasons, they do
not decay; it has also been shown that the rate of pair
annihilation cannot reduce the monopole number sufficiently
\cite{weinberg}.  Their abundance is expected to be too high to be
consistent with astrophysical observations of magnetic fields in
our Galaxy \cite{parker} and consequences of catalyzed nucleon
decay in stars \cite{catalysis,cat2}.  This overabundance of GUT
monopoles is known as the monopole problem.

An inflationary epoch \cite{guth} may reduce the density of
monopoles in the Universe (for some alternative approaches see
e.g. \cite{alt}).  The rapid expansion phase dilutes the monopole
abundance. The re-heating temperature must be sufficiently low in
order not to produce the monopoles again. On the other hand, since
any initial baryon-antibaryon asymmetry will be also washed out,
the re-heating temperature must be sufficiently high in order to
provide a mechanism for the observed baryon-antibaryon asymmetry.
Here we explore new solutions to the monopole problem which, for
example, would obviate the need for a low re-heating temperature.

As the universe expands, the cosmological horizon grows to
encompass many monopoles.  The predicted initial ratio
\cite{VilenkinShellard} between the monopole and entropy density
is given by
\be \label{initial} \left( \frac{n_m}{s} \right)_i \sim p
g_*^{1/2}(\eta/M_{Pl} )^3 \sim 10^{-12}\, , \ee where $p \sim 0.1$ is
a geometrical factor and $g_* \sim 100$ is
the number of effectively massless degrees of freedom at a given
temperature. This ratio would remain constant as the universe
adiabatically expands.  However non-adiabatic processes, such as
monopole/antimonopole annihilation or accretion by black holes, reduce this ratio. 

Here we first review the relic monopole abundance subsequent
to annihilation effects only, and later include a new effect: the fact
that primordial black holes can eat monopoles.  The relic abundance
is far too large with annihilation only, but this problem can be
resolved with black holes.

The motion of a
monopole in a plasma is basically a random walk where the typical step
is the mean free path. During each step a monopole moves with thermal
velocity $v_T = \left( 2T/M_m \right)^{1/2}$. The mean free path $l$
depends on the scattering of a monopole on charged particles in plasma and
is given by \cite{VilenkinShellard}:

\be
l \approx \frac{M_m^{1/2}}{bT^{3/2}} \, .
\ee
The factor
$b$ depends on the number of degrees of freedom of light charged
particles and its value is $b \sim 100$.
Monopole-antimonopole capture occurs when $g^2/r \sim T$, which gives
the capture radius $r_c \sim g^2/T$. Annihilation effectively
takes place as long as the mean free path of a monopole is smaller than
the capture radius for annihilation \cite{ZKP}. This implies that the
annihilation practically stops at the final temperature

\be T_f^{\rm ann} \sim \frac{M_m}{g^4b} \sim 10^{11} {\rm GeV} \,
, \ee
 and the monopole to entropy density ratio freezes
at the final value \cite{weinberg,VilenkinShellard}

\be \label{nsf} \left( \frac{n_m}{s} \right)_f \sim 1/(g^6 b
g_*^{1/2}) M_m/M_{Pl} \sim 10^{-10}\, . \ee

The annihilation mechanism is effective only if the initial
monopole to entropy density ratio is greater than the one given by
(\ref{nsf}). Otherwise, the initial ratio remains constant as the
universe expands. For our fiducial choice of parameters with
$M_m \sim 10^{16}$ GeV, annihilation is irrelevant and the initial
 monopole
to entropy density ratio of Eq.(\ref{initial}) remains constant;
for other parameters and monopole masses, however, Eq.(\ref{nsf}) should
be used.
The entropy density at a given temperature $T$
is $s \sim g_* T^3$.  The total mass of
 monopoles
within the Hubble volume is thus
\be
\label{mtot}
M_{\rm tot} \sim \left(
  \frac{n_m}{s} \right)_f s V_h M_m \approx
10^{-2}\left(\frac{n_m}{s}\right)_i \left(
\frac{M_{Pl}}{T}\right)^3  M_m
 \, , \ee
where $V_h \sim \frac{4\pi}{3}\tau_h^3$ is a Hubble volume at a
given Hubble time $\tau_h = 0.3 g_*^{-1/2} M_{Pl}/T^2$.  This mass
will equal the total mass within the Hubble volume ($M_h \sim 0.3
g_*^{-1/2} M_{Pl}^3/T^2$) for $T \sim 10^3$GeV. Thus, monopoles
dominate the energy density of the Universe at $T < 10^3$GeV. Thus the
predicted present day monopole flux 
grossly violate many
late-time cosmological constraints, if one considers annihilation
as the only mechanism for reducing the monopole number.

Now we will consider the effects of black holes in reducing the
monopole number.  The early universe can produce large numbers of
primordial black holes, via a number of processes
\cite{pbh1,pbh3,pbh5,pbh}.  The earliest mechanism for black hole
production can be fluctuations in the space-time metric at the Planck
epoch. Large number of primordial black holes can also be produced by
nonlinear density fluctuations due to oscillations of some (scalar)
field.  If within some region of space density fluctuations are large,
so that the gravitational force overcomes the pressure, we can expect
the whole region to collapse and form a black hole. Black holes can
also be produced at first and second order phase transitions in the
early universe \cite{pbh5}. Collapse of cosmic string loops and closed
domain walls also yields black holes.

We will show that black holes could solve the monopole problem by
capturing them before they become dangerous for the standard
cosmology. Since the black holes evaporate, much of the dangerous excess monopole energy density
would then be converted into harmless radiation.  If the universe
contains enough black holes to remove monopoles quickly enough, but
not so many of them to cause deviations from standard cosmology, then
the monopole problem would be solved.

Primordial black holes can have a variety of masses. A typical
mass would be that of the mass inside the horizon at the time of
formation, so that the mass would range roughly from $M_{Pl}$ (black
holes formed at the Planck epoch) to $M_{sun}$ (black holes formed at
the QCD phase transition).  However,
the mass could be much smaller\cite{pbh3}.  For example,
black holes formed by collapsing closed domain walls and string loops
are typically much lighter than a horizon mass. After formation black
holes can evolve further by evaporating, merging with each other and
accreting surrounding matter. Since the number of black holes of a
given mass within a horizon volume as a function of time is strongly
model dependent, we do not limit ourselves to any particular model. We
instead solve the dynamical equation for the time evolution of the
number density of monopoles with free parameters that determine the
abundance of black holes, and then estimate the range of these
parameters that is required for the solution of the monopole
problem. We will show that black holes of mass $M_{bh} < 10^{9}$gm (a
mass range that does not usually put any serious constraints on
standard cosmology) are dynamically capable of removing the monopoles.

Monopoles move with non-relativistic velocities $v_M<<c$ in the
early universe, since they are magnetically charged and their
velocity is damped due to interaction with surrounding plasma. The
cross section for gravitational capture of a non-relativistic
massive particle (monopole) by a black hole of gravitational
radius $R_{bh}$ is \cite{FrNo}: \be \label{cs} \sigma_g = 4 \pi
(c/v_M)^2 R_{bh}^2 . \ee The monopole flux is given by $n_m v_m$,
where $n_m$ is the number density. If there is only one black hole
in the Hubble volume, the number of monopoles that get captured by
the black hole, during the Hubble time $\tau_h$, is $n_m v_m
\sigma_g \tau_h$. The total number of captured monopoles is
obtained by multiplying this quantity by the number of black holes
within the Hubble volume, $n_{bh} V_h$, where $V_h$ is the Hubble
volume.  This number should be compared with the total number of
monopoles contained within the Hubble volume, $n_m V_h$, thus
giving the condition
\begin{equation}
\label{mc}
n_{bh} \sigma_g  v_m  \tau_h  > 1 \, ,
\end{equation}
This inequality represents the condition under which the
characteristic time for monopole capture is shorter than the Hubble
time. We will now ascertain the resulting number density of monopoles
by examining the dynamics in more detail.
As we will see, the term on the left hand side 
of (\ref{mc}) will play the
role of the suppression factor for the evolution of the monopole
number density. How big this factor should be in order to solve the
monopole problem, we will see from the equation that describes the
dynamics of the process.

The dynamical equation (i.e. the detailed balance equation) for the
time evolution of number density of monopoles in comoving volume
is

\be \label{be} \frac{dn_m}{dt}=-A n_m^2 -n_{bh} \sigma_g v_m n_m
-3\frac{{\dot a}}{a} n_m \, . \ee The first term on the right hand
side is present as long as the monopole-antimonopole annihilation is
taking place. The parameter $A$
was estimated in \cite{ZKP} as $A = g^2/(bT^2) =
g^2g_*^{1/2}t/(0.3bM_{Pl}) \approx 10t/M_{Pl}$.
The second term takes into account the monopole capture by
the black holes while the third term is due to expansion of the
universe ($a$ is the scale factor of the universe). 

We now examine the term $n_{bh} \sigma_g v_m$. The energy density of
the black holes at any given time must be less than the energy density
of the universe, so that $\rho_{bh} = M_{bh} n_{bh} \sim f g_* T^4$,
where $f \leq 1$, while $M_{bh}$ is the mass of a typical black hole
at that scale. The number density of black holes is thus $n_{bh} \sim
fg_* T^4/M_{bh}$. For a given black hole mass $M_{bh}$, the radius
$R_{bh} = 2 M_{bh}/M_{Pl}^2$. The relative black hole-monopole
velocity is just a monopole random walk velocity $v_m = v_T/\sqrt{N}$,
where $v_T$ is a thermal velocity while $N$ is the number of random
walk steps. At some distance $r$, we have $\sqrt{N} =r/l$, where $l$
is the length of the step (a mean free path). The characteristic
distance is the gravitational capture radius calculated from the
condition $M_{bh}M_m/(M_{Pl}^2r_c) \sim T$. This gives $v_M = \sqrt{2}
M_{Pl}^2/(bM_{bh}M_m)$. The mass of a typical black hole is just a
fraction of the mass within the horizon at any given time. In other
words, 
\be
\label{defnbeta}
M_{bh} \sim \beta \, 0.3 \, g_*^{-1/2} \frac{M_{Pl}^3}{T^2} ,
\ee
where $\beta \leq 1$. In the absence of complete knowledge of the
distribution of black holes as a function of time, an assumption must
be made about the time dependence of $f$ and $\beta$.  
In our analysis of the dynamics, we will assume that the parameters 
$f$ and $\beta$ do
not change significantly with time as they describe some average features
of the system\footnote{We note that this assumption requires
black holes to form at a range of temperatures or to grow
due to accretion.}.  
Substituting these relations into $n_{bh} \sigma_g v_m$ we obtain

\be n_{bh} \sigma_g  v_m = f \beta^2 b M_m \approx 0.1 f \beta^2
M_{Pl} \,. \ee

The dynamical equation (\ref{be}) now reads

\be \frac{dn_m}{dt}=-\frac{10t}{M_{Pl}} n_m^2 - 0.1 f \beta^2
M_{Pl}n_m -\frac{3}{2t} n_m \, . \ee

The solution to this equation is

\be \label{be1} n(t) = \frac{e^{-0.1 f \beta^2 t/t_{Pl} }}{ C_1
t^{3/2} +
 (0.1\pi/M_{Pl}^3f\beta^2)^{1/2}
t^{3/2} {\rm erf} \left( \sqrt{0.1 f \beta^2 t}    \right)  } \, ,
\ee where $ {\rm erf} (x) = 2/\sqrt{\pi} \int_0^x e^{-t^2}dx$ and
$C_1$ is a constant of integration.

The gravitational capture of monopoles by the black holes effectively
takes place as long as the capture radius is larger than the mean free
path of monopoles. This condition is satisfied for
$T<(\beta^2g_*^{-1}b^2M_m)^{1/3}M_{Pl} \approx (0.1 \beta^2)^{1/3}
M_{Pl}$. While the electromagnetic monopole-antimonopole capture stops
at some finite temperature, the gravitational capture starts at some
high temperature and practically never ends. This is the consequence
of the fact that the universe can accommodate larger and larger black
holes as time passes, so that the gravitational capture radius grows
faster than the mean free path of monopoles. Hence we can neglect the
monopole annihilation contribution in eq. (\ref{be1}) (the second term in
denominator). We note that in general 
dropping the second term is a conservative
assumption in the sense that we are underestimating the destruction of
monopoles. Thus we have

\be \label{nsp} \frac{n(t)}{s} =\left( \frac{n(t)}{s} \right)_i
e^{-0.1 f \beta^2 (t-t_i)/t_{Pl} } \, , \ee
where the initial
ratio between the monopole and entropy density is given by eq.
(\ref{initial}). We see that the gravitational capture of the
monopoles by the black holes exponentially reduces this ratio. We
note that the most massive black holes contribute the most, while
the contribution of much lighter ones can be neglected. Also, if
$t_i \ll t$ we can neglect the term $t_i$ in the exponential. This
will be true in practically all the cases of interest here.

Now, we can derive the values of $f$ and $\beta$ needed for a
solution of the monopole problem. The first bound on the monopole
abundance comes from the requirement that the universe must be
radiation dominated at nucleosynthesis, i.e. $T \sim $MeV.
This implies that the ratio between the monopole and entropy
density can be at most

\be \label{mev} \left( \frac{n}{s} \right)_{T ={\rm MeV}} \leq
\frac{1{\rm MeV}}{M_m} = 10^{-19} \, . \ee

We can derive a more stringent bound by requiring that the mass
density of the monopoles today should not exceed the critical
density: $M_m n_m < \rho_c$. This yields

\be \label{today} \left( \frac{n}{s} \right)_{{\rm today}} \leq
\frac{\rho_c }{M_m s_0} = 10^{-24} \, , \ee where we used the
current value of the entropy density  $s_0 =10^3 {\rm cm}^{-3}$.
There are a number of other late time constraints \cite{parker}.
For example, since monopoles are magnetically charged, they can be
accelerated by galactic magnetic fields. In this process the
galactic  magnetic field gets dissipated. In order to avoid a
conflict with observations, the upper limit on monopole number
density is

\be n_m < 10^{-20} {\rm cm}^{-3} \, . \ee

Certainly, the strongest constraint  comes from the consequences of
monopole induced baryon catalysis in neutron stars \cite{cat2}:

\be n_m < 10^{-25} {\rm cm}^{-3} \, . \ee
This implies the bound
on a current ratio between the monopole and entropy density:

\be \label{nsb}
\left( \frac{n}{s} \right)_{\rm today}  < 10^{-28} \, . \ee

Substituting eq. (\ref{initial}) and  eq. (\ref{nsb}) into eq. (\ref{nsp})
we can derive

\be \label{nc} f \beta^2 t_f/t_{Pl} \ge 100 \, , \ee 
or, equivalently,
\be \label{tempf}
f \beta^2 \left({m_{pl}/T_f}\right)^2 \ge 100 
\ee
for our fiducial value of $M_m = 10^{16}$ GeV.
This is the
necessary condition for a solution of the monopole problem. Here,
$t_f$ and $T_f$ denote time and temperature respectively
when the process of gravitational capture of
monopoles by black holes stops. 
We now discuss whether or not the universe can
reasonably produce black holes that satisfy this condition.

Black holes more massive than $10^{15}$ gm have lifetimes longer than
the present age of the universe. While these black holes are
dynamically capable of swallowing the monopoles, the energy density
that was in the monopoles remains in the form of mass rather than
radiation and hence overcloses the universe.  Hence we must consider
lighter black holes.

Black holes of mass $M_{bh} < 10^{15}$gm
evaporate before the present epoch, thereby converting the
monopole energy density to radiation which redshifts away.  However,
observational constraints from earlier epochs have been previously
examined by other authors\cite{pbh} and found to be more severe. These
constraints are sensitive to the assumptions in the model, e.g., model
of formation of the black holes, equations of state in different
epochs of the early universe, whether or not the universe experiences
a period (or multiple periods) of inflation, etc.  Ref.  \cite{pbh}
shows that black holes with mass over $10^{9}$g could not have been
produced with appreciable abundance relative to the total energy
density without violating observational constraints due to the cosmic
microwave background (CMB), nucleosynthesis, geometry of the universe,
etc.  On the other hand, black holes less massive than $10^9$gm are
not seriously constrained by observations: they evaporate completely
within $0.1$sec, i.e. before nucleosynthesis takes place, and are
compatible with results of CMB experiments.

Thus, we work with the constraint that only black holes of $M_{bh}
\leq 10^{9}$gm can be responsible for removing monopoles from the
universe. 
Using Eqs.(\ref{tempf}) and (\ref{defnbeta}), together with
this lower bound on the black hole mass, we find
\be
\label{minTf}
T_f > {3 \times 10^{-13} M_{pl} \over g_*^{1/2} f^{1/2}}
= 3 \times 10^6 {\rm GeV} \left({f \over 0.01}\right)^{-1/2}
\left({g_* \over 100} \right)^{-1/2} .
\ee

\begin{figure}[tbp]
\centerline{\epsfxsize = 0.95 \hsize \epsfbox{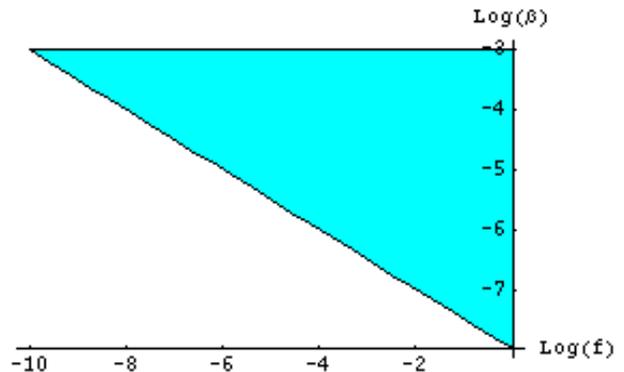}}
\caption{The allowed parameter range for the solution of the
monopole problem (for $M_m = 10^{16}$GeV) lies above the curve $\beta =
(10^{-16}/f)^{1/2}$ (shaded region on the Log-Log plot). Parameter $f$
is the fraction of the total energy density of the universe that must be
in black holes, while $\beta$ is the fraction of the horizon mass  that
represents a typical (average) black hole at that time. Even for values of
$f \ll 1$ we can have reasonable values for $\beta$.} \label{fbeta}
\end{figure}

As an example, we will consider $T_f = 10^9$GeV, or, 
equivalently $t_f \sim 10^{18} t_{Pl}$, for the final
temperature. Then the condition (\ref{nc}) becomes 
\be f\beta^2 \geq
10^{-16} \, . \ee 
We plot the allowed range in Fig. ~\ref{fbeta}.

For example, if $f \sim 10^{-2}$ ($1 \%$ of the energy density of
the universe is in black holes), the allowed range for $\beta$ is
$\beta \geq 10^{-7}$, which means that the mass of a typical black
hole at that time should be at least $10^{-7}$ of the horizon
mass. This is not difficult to envision. The horizon mass at $T
\sim 10^{9}$GeV  is roughly $M_h \sim 10^{13}$gm, so the average
black holes mass should be $M_{bh} \geq 10^{6}$gm. Note that the
number of black holes within a horizon is roughly given by $N_{bh}
= f/\beta$, but since $\beta$ gives only an average description of
the system, the real number of black holes could be quite  different
from $N_{bh}$ defined in this way. We therefore do not use
$N_{bh}$ as a reliable quantifier.

We have shown that, as far as the dynamics of the process is
concerned, a reasonable
number of primordial black holes is capable of capturing monopoles
from the early universe. We now discuss problems that might
potentially arise.

After absorbing monopoles, a black hole will become charged. In our
calculation we assumed that the monopole capture by the black hole is
charge blind, i.e. we neglected the electromagnetic interaction between
the monopole and black hole. Now, we justify this assumption.  Let's
consider a black hole which has already captured $N_m$ monopoles and
antimonopoles.  Most of the monopoles and
antimonopoles inside the black hole will annihilate, leaving
$\sqrt{N_m}$ monopoles inside (due to magnetic charge fluctuations).
The gravitational mass of the black hole is enhanced to $M_{bh} + N_m
M_m$; the magnetic charge of the black hole becomes $\sqrt{N_m}g$, where
$g$ is a magnetic charge of the monopole.  The gravitational force
between this black hole and a monopole outside it is then 
\be F_g = G
(M_{bh} + N_m M_m) M_m /r^2 \, . \ee 
The corresponding magnetic force is
\be F_m = \sqrt{N_m} g^2 /r^2 \, . \ee 
If $N_m$ is large ($N_m> 10^{15}$),
gravitational force is always greater than the magnetic. Even if $N_m$
is small, one can easily check that the gravitational attraction
between a monopole of mass $10^{16}$GeV (with a unit charge $g$) and a
black hole that captured $N_m$ monopoles is larger than the
electromagnetic repulsion between them for any black hole with
\begin{equation}
\label{eq:magforce} M_{bh} > 10^4 \sqrt{N_m} M_{Pl} ,
\end{equation}
a condition which is easily satisfied for the black holes of
interest.

Another problem that might arise in this scenario can come from
magnetic charge fluctuations: a black hole that captures $N_m$
monopoles and anti-monopoles will typically have accumulated a
residual net magnetic charge of $\sqrt{N_m}$. The question is, what
happens to this magnetic charge? One possibility would be that
black holes Hawking-radiate monopoles. A monopole is a highly
coherent state of many gauge quanta and emission of a monopole by
a black hole will be highly suppressed. Even if this process is
somehow allowed \cite{page}, the Hawking temperature of a black
hole becomes of order of a monopole mass only at the end of
 evaporation, and a black hole could radiate much less
monopoles than originally eaten. Generically, since the Hawking radiation 
can not violate the gauge symmetry, a black  hole can not  evaporate
completely. Instead it leaves a remnant --- an extreme magnetically
charged Reissner-Nordstrom black hole. The mass of the  (non-rotating)
remnant $M_r$ must be greater than the magnetic charge $Q_m$ of the black
holes, $M_r \geq Q_m$ (in appropriate units), or otherwise the remnant
would be a naked singularity. One has to check whether these remnants
violate some of the observational constraints. Due to the attractive force
between the remnants with opposite charges, their number density
will be reduced (charges annihilate and the rest of the mass
evaporates freely). However, here we adopt the most
conservative  scenario in which all the black holes which
participate in accreting monopoles leave massive remnants. 
The mass density
$\sqrt{N_m}$ that remains trapped in the extremal
Reissner-Nordstrom black holes could potentially
be problematic.  The energy density in these remnants
must be less than that of ordinary radiation at the time
of primordial nucleosynthesis, in order for predictions
of element abundances to be unaffected.
Here we will compare the residual number of monopoles inside black
holes and ensure that it is smaller than the number of monopoles
allowed by the nucleosynthesis bound in Eq.(\ref{mev}).
The total number of monopoles within the horizon
at temperature $T$ is $N_{tot} = M_{\rm tot}/M_m$, or, using
Eq.(\ref{mtot}),
\be \label{nm} N_{tot} \sim 10^{-2}\frac{n_m}{s} \left(
\frac{M_{Pl}}{T}\right)^3
 \,  .\ee
We note that this number is proportional to $n_m/s$.
If all of the monopoles within horizon are
eaten by black holes at some temperature (e.q. $T \sim 10^{9}$GeV), then
 the remnant monopole mass inside the black holes is $\sqrt{N_m} \propto
\sqrt{\left({n_m \over s}\right)_i}$ given in Eq.(\ref{initial}). On the
other hand, the allowed number of monopoles is proportional to $N_{\rm
allowed} \propto (n_m/s)_{\rm MeV}$ given in Eq.(\ref{mev}). Thus, we can calculate
that at $T \sim 10^{9}$GeV, there are $N_m \sim 10^{16}$
monopoles inside a horizon volume.
Practically all of them are eaten by the black holes.
Mass density of $\sqrt{N_m} \sim 10^{8}$ monopoles can not be
eliminated (under our conservative assumptions). However, the
number of monopoles that the universe can tolerate is $N_m \sim
10^{9}$. Thus, the mass density of the remnants
does not violate observational constraints.

Later, there can be further annihilation of monopole/
antimonopole pairs inside galaxies and clusters of galaxies
where these objects can clump; hence we shy away from extrapolating
the freeze-out density in Eq.(\ref{initial}) (or  Eq.(\ref{nsf})) all the
way to the current epoch to estimate the mass density of remnants today
and then applying the constraint of Eq(\ref{today}).

In conclusion, we have presented a possible solution to the
cosmological monopole problem in which primordial black holes
efficiently eliminate the magnetic monopoles from our universe.
Under very conservative assumptions,
black holes of mass
 \be M_{bh} < 10^{9} {\rm gm}  \ee are dynamically capable of
solving the cosmological monopole problem, while their required
abundance does not violate any of the observational constraints.  We
did not assume any particular model for primordial black holes
formation. Instead, we derived the bounds on abundance of black
holes that can reasonably exist in our universe and are capable of
removing monopoles. Black holes have a large cross-section for
capturing non-relativistic monopoles and the parameter space for
solving the monopole problem is not very restrictive. 
In a similar manner, black holes can in principle
remove other unwanted massive relics from the early universe \cite{dwp}.

\vspace{12pt} {\bf Acknowledgments}:\ \
The authors are grateful to Josh Frieman, Glenn Starkman, Tanmay
Vachaspati and especially Fred Adams for very useful discussions.
We thank the DOE and the Michigan Center for Theoretical Physics for
support at the University of Michigan.

\end{document}